\documentclass{PoS}

\newcommand{\Xmax}{X_{\mathrm{max}}}
\newcommand{\mrd}{\mathrm{d}}
\newcommand{\dEdX}{\frac{\mrd E}{\mrd X}}
\newcommand{\Nch}{N_{\mathrm{ch}}}

\title{Description of longitudinal profiles of showers dominated by Cherenkov light}

\ShortTitle{Description of longitudinal profiles of showers dominated by Cherenkov light}

\author{\speaker{Vladim\'ir Novotn\'y}\\
        Charles University, Faculty of Mathematics and Physics, Institute of particle and Nuclear Physics, Prague, Czech Republic\\
        E-mail: \email{novotnyv@ipnp.troja.mff.cuni.cz}}

\author{Dalibor Nosek\\
        Charles University, Faculty of Mathematics and Physics, Institute of particle and Nuclear Physics, Prague, Czech Republic\\
        E-mail: \email{nosek@ipnp.troja.mff.cuni.cz}}

\abstract{With the aim to describe the longitudinal development of Cherenkov dominated showers we investigate the energy deposit and the number of charged particles in air showers induced by energetic cosmic rays. Based on the Monte Carlo simulations, discrepancies between different estimates of calorimetric energies are documented. We focus on the energy deposit profiles of air showers deducible from the fluorescence and Cherenkov light generated along CONEX and CORSIKA cascades.}

\FullConference{35th International Cosmic Ray Conference -- ICRC2017\\
		10-20 July, 2017\\
		Bexco, Busan, Korea}

\begin{document}

\section{Introduction}

One of the most interesting observables that is measured by cosmic ray experiments is the longitudinal profile of air showers induced by primary particles with very high energies.
This profile is determined by the energy deposit along the shower evolution, 
when expressed as a function of the shower depth in the Earth atmosphere, 
i.e. $\left( \dEdX \right)(X)$.
The energy deposit is usually determined by the measurement of the fluorescence 
light produced in the atmosphere by the excitation of nitrogen molecules.
Induced by this excitation, the intensity of registered fluorescence light 
is directly proportional to the value of deposited energy $\left( \dEdX \right)(X)$.

Besides the fluorescence light, also the Cherenkov light is generated 
during the shower development. 
Its production is proportional to the number of charged particles 
having energies over the Cherenkov threshold in the atmosphere.
Consequently, the Cherenkov component is registered together with 
the fluorescence light by means of the fluorescence detectors.

For the shower profile reconstruction, an entanglement between 
the fluorescence and Cherenkov light is usually modelled by using the Cherenkov-Fluorescence matrix, see Ref.\cite{CFM}.
Detailed parametrizations of the number of charged particles in 
air showers as well as the profiles of energy deposit are presented in Ref.\cite{nerling}.
These parametrizations were obtained using simulations produced 
within the CORSIKA software \cite{corsika}.

An essential ingredient of the air shower reconstruction is the
mean ionisation loss rate $\alpha(X)$ that relates the energy deposit, $\left( \dEdX \right)(X)$, and the number of charged particles, $\Nch(X)$, in the shower at a given depth in the atmosphere $X$
\begin{equation}
\label{eq:alpha}
\alpha (X) = \frac{1}{\Nch(X)} 
\left( \dEdX \right)(X).
\end{equation}
This function is usually expressed in terms of the shower age $s$ as
\begin{equation}
s(X) = \frac{3}{1+2\Xmax/X},
\end{equation}
and where $\Xmax$ denotes the maximum depth of the energy deposit profile
of a shower.
Relying upon CORSIKA cascades, the parametrization of $\alpha(s)$
was given in Ref.\cite{nerling}
\begin{equation}
\label{eq:old_par}
\alpha(s) = \frac{c_1}{(c_2+s)^{c_3}} + c_4 + c_5 \cdot s,
\end{equation}
where 
$c_1 = 3.90883~\mathrm{MeV~g^{-1}~cm^{2}}$, $c_2 = 1.05301$, $c_3 = 9.91717$, $c_4 = 2.41715$ $\mathrm{MeV~g^{-1}~cm^{2}}$, $c_5 = 0.13180~\mathrm{MeV~g^{-1}~cm^{2}}$.
These values are valid for an energy cut on the electromagnetic component at $E_{\mathrm{cut}}=1$~MeV and cut for hadronic and muonic component $E_{\mathrm{had-cut}}=100$~MeV. The low--energy interaction model Geisha 2002 \cite{geisha} together with high--energy interaction model QGSJET 01 \cite{qgsjet} were used.

The aim of this work is to check the impact of the parametrization of $\alpha(s)$ to the calorimetric energy determination in the case that Cherenkov light dominates the measured light flux.
Is such a case, the light is emitted according to the $\Nch(X)$ profile and $\alpha(s)$ is used to get the $\left( \dEdX \right)(X)$ profile from collected light flux.

Events measured by fluorescence telescopes that are dominated by Cherenkov light have lower detection energy threshold then those dominated by fluorescence light.
Because of that, we are interested mainly in energies below $10^{17}$~eV, i.e. we chose EPOS LHC generated $10^{16}$~eV proton primaries for all studies presented in this work.

A study of precise Cherenkov emission model is beyond the scope of this work and could be found, for example, in Refs.\cite{nerling, cherenkov}.

Showers generated by CONEX \cite{conex} and CORSIKA \cite{corsika} tools are investigated.
The method of the calorimetric energy estimation is given in Section \ref{sec:method} and results of the simulation studies are summarized in Section \ref{sec:results}.

\section{Method}
\label{sec:method}
To mimic a real air shower reconstruction procedure, we adopt the following method of estimation of calorimetric and reconstructed energy.
In our calculations, the calorimetric energy of a shower,
$E_{\mathrm{cal}}$, is given by the sum of individual energy deposits
in the simulated profile bins, d$E_i$, corrected for ground effects, i.e.
\begin{equation}
E_{\mathrm{cal}} = 
\sum_{i=0}^n \mrd E_i + E_{\mathrm{ground-EM}} + 
0.61 E_{\mathrm{ground-had}}.
\end{equation}
Here, the factor of $0.61$, that modifies the energy of the hadronic
component at ground, comes from results of Ref.\cite{barbosa}.
It accounts for the fraction of hadronic energy that would turn 
into the electromagnetic component if the shower development continued under ground.
Calculations of $\mrd E_i$ differ from CORSIKA to CONEX generators.
In the CONEX case, the $\mrd E_i$ is calculated when solving cascade equations and is directly accessible from the CONEX output file.
In the case of CORSIKA, the energy deposits of different particle species are listed in the simulation output.
The $\mrd E_i$ is then given by
\begin{equation}
\mrd E_i = \mrd E_{\mathrm{tot}} - 0.575 \mrd E_{\mathrm{\mu-cut}}
 - \mrd E_{\mathrm{\nu}},
\end{equation}
where $\mrd E_{\mathrm{tot}}$ is the total energy deposited by all particle types. $\mrd E_{\mathrm{\nu}}$ is the energy of created neutrinos and $\mrd E_{\mathrm{\mu-cut}}$ 
is the energy of muons
that fall under the $E_{\mathrm{had-cut}}$.
The factors are determined 
from Refs.\cite{barbosa, pierog}.

In order to estimate reconstructed energies of showers, the energy deposit profile is described by a Gaisser-Hillas (GH) function written
\begin{equation}
\left( \dEdX \right)_{\mathrm{GH}}(X) = 
\left( \dEdX \right)_{\mathrm{max}} 
\left(\frac{X-X_0}{\Xmax-X_0}\right)^\frac{\Xmax-X_0}{\lambda} 
\exp\left(\frac{\Xmax-X}{\lambda}\right),
\end{equation}
where $\left( \dEdX \right)_{\mathrm{max}}$, $\Xmax$, $X_0$ and $\lambda$ are parameters to be fitted.
Knowing this profile directly from simulations or estimating it 
from the number of charged particles along the shower evolution 
using the $\alpha$-function, we introduce the reconstructed calorimetric 
energy, $E_{\mathrm{GH-fit}}$. 
This energy is given by the analytic formula 
\begin{eqnarray}
E_{\mathrm{GH-fit}} = 
\int\limits_0^{\infty} \left( \dEdX \right)_{\mathrm{GH}}(X) \, \mathrm{d}X = 
\left( \dEdX \right)_{\mathrm{max}} e^Y Y^{-Y} \Gamma \left(Y+1\right),
\end{eqnarray}
where $Y = \frac{\Xmax-X_0}{\lambda}$.

\begin{figure}
\begin{center}
\includegraphics[scale=0.65]{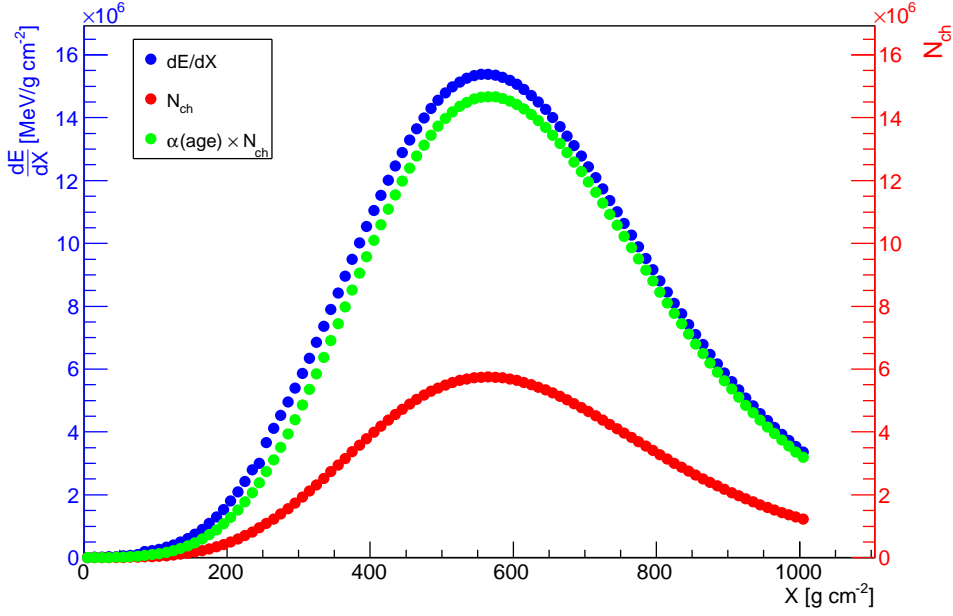}
\end{center}
\caption{Illustration of the discrepancy between $\left( \dEdX \right)(X)$ and $\alpha \cdot \Nch(X)$ profiles. Randomly selected proton shower with the energy of $10^{16}$~eV generated by CONEX v4r37 is shown.}
\label{fig:single}
\end{figure}

The impact of the $\alpha(s)$ function on the reconstructed energy is quantified by the difference between $E_{\mathrm{GH-fit}}$ of $\left( \dEdX \right)(X)$ and $\alpha \cdot \Nch(X)$ profiles.
An illustration of the discrepancy between these profiles calculated with the use of CONEX v4r37 is depicted in Fig.\ref{fig:single}.

\section{Results}
\label{sec:results}

\begin{figure}
\begin{center}
\includegraphics[scale=0.65]{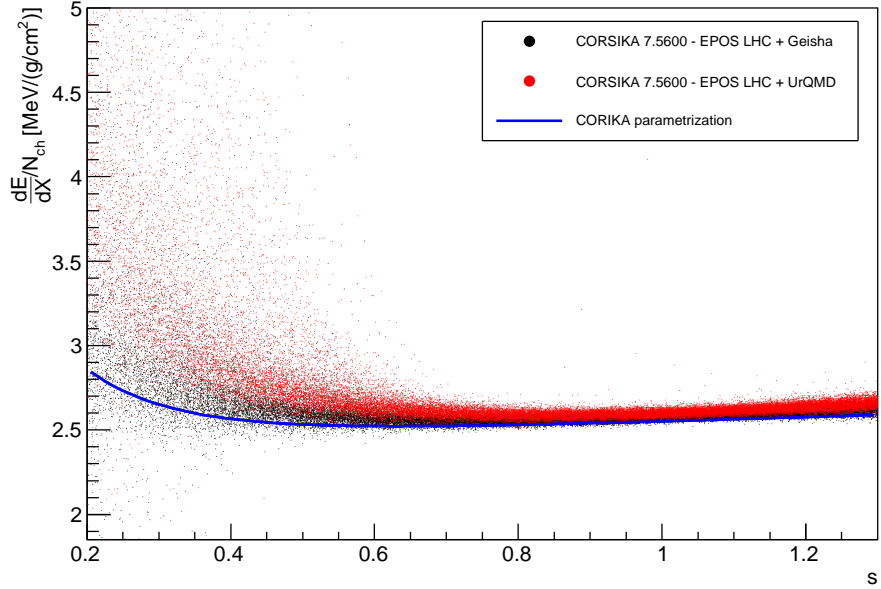}
\end{center}
\caption{Ionization loss rates calculated with the use of CORSIKA full MC simulation tool. High--energy interaction model EPOS LHC together with Geisha 2002 (black) and UrQMD 1.3 (red) low--energy interaction models were used. Parametrization of $\alpha$ according to Eq.(\ref{eq:old_par}) is shown in blue. 500 showers were generated for each model. Points represent values for individual showers.}
\label{fig:scat_corsika}
\end{figure}

\begin{figure}
\begin{center}
\includegraphics[scale=0.65]{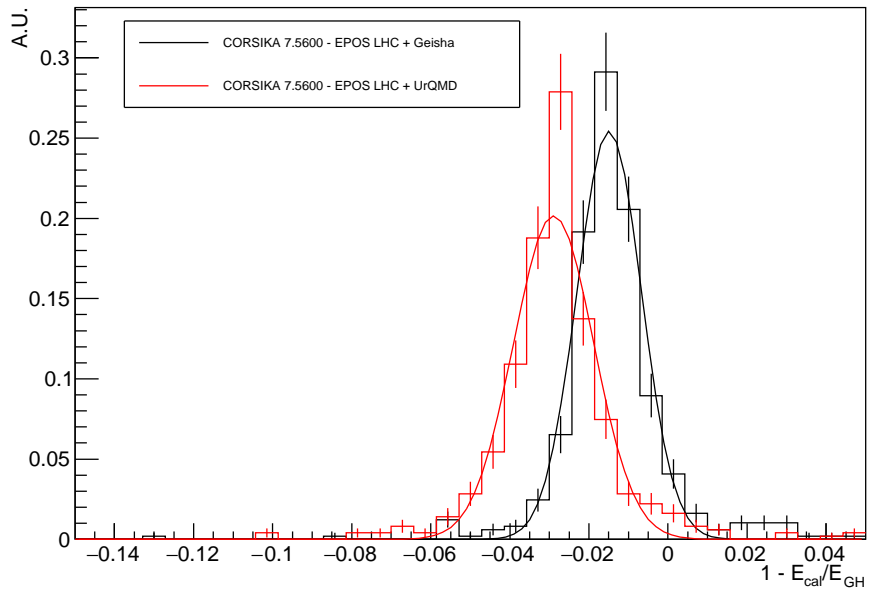}
\end{center}
\caption{Discrepancy between estimates of simulated and reconstructed calorimetric energies. Colour scheme is the same as in Fig.\ref{fig:scat_corsika}. For details see the text.}
\label{fig:dE_corsika}
\end{figure}

At first, we checked if the parametrization given in Eq.(\ref{eq:old_par}) reasonably describes the effective ionization loss rate determined by more recent CORSIKA version 7.5600\footnote{A thinning level of $10^{-6}$ was used in our calculations.} than the one that was used for the original calculation in Ref.\cite{nerling}.
Obtained results are shown in Fig.\ref{fig:scat_corsika} where we document a slight discrepancy between results of Geisha 2002\footnote{The parametrization given in Eq.(\ref{eq:old_par}) was calculated with the use of this model.}~\cite{geisha} and UrQMD 1.3\footnote{The minimal available $E_{\mathrm{had-cut}}$ for this model is 0.3 GeV. We used $E_{\mathrm{had-cut}}=1$~GeV which corresponds to the CONEX default settings. Part of the discrepancy has to be attributed to such setting.}~\cite{urqmd} low--energy interaction models.
The total impact of the $\alpha$ parametrization and GH--fit procedure, as described in Section \ref{sec:method}, to the bias in reconstructed energy is of $1.2\%$ and $2.5\%$ for Geisha and UrQMD models, respectively.
Standard deviations of the histograms, which correspond to the shower to shower fluctuations, are roughly $1\%$.
The net effect on reconstructed energy is visualized in Fig.\ref{fig:dE_corsika}.

\begin{figure}
\begin{center}
\includegraphics[scale=0.65]{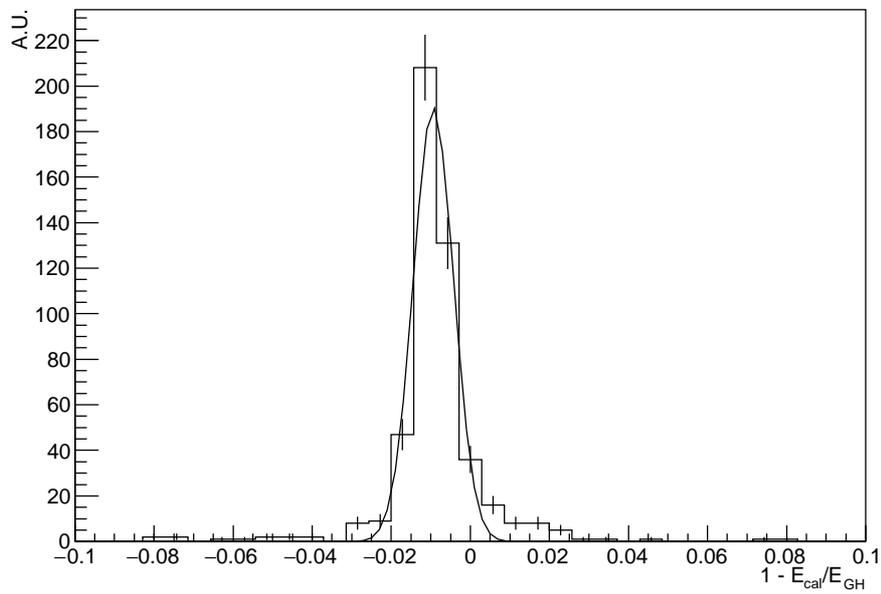}
\end{center}
\caption{Impact of the GH--fit procedure to the estimate of the reconstructed calorimetric energy. CONEX generated showers were used to produce the plot.}
\label{fig:dE_GH}
\end{figure}

Because of the bias caused by GH--fit procedure of $1\%$, calculated for CONEX showers and shown in Fig.\ref{fig:dE_GH}, we can conclude that the impact of $\alpha$ parametrization to the energy reconstruction is very small if we consider full Monte Carlo (MC) simulated showers generated by CORSIKA.

\begin{figure}
\begin{center}
\includegraphics[scale=0.65]{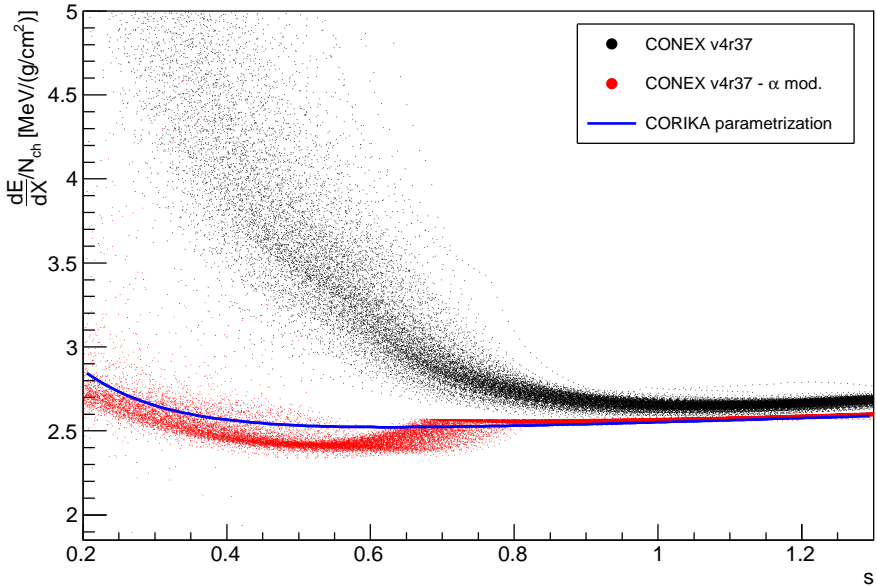}
\end{center}
\caption{Ionisation loss rates calculated with the use of CONEX generated showers. Black dots represent the CONEX v4r37 results and red dots correspond to the modification described in the text.}
\label{fig:scat_conex}
\end{figure}

\begin{figure}
\begin{center}
\includegraphics[scale=0.65]{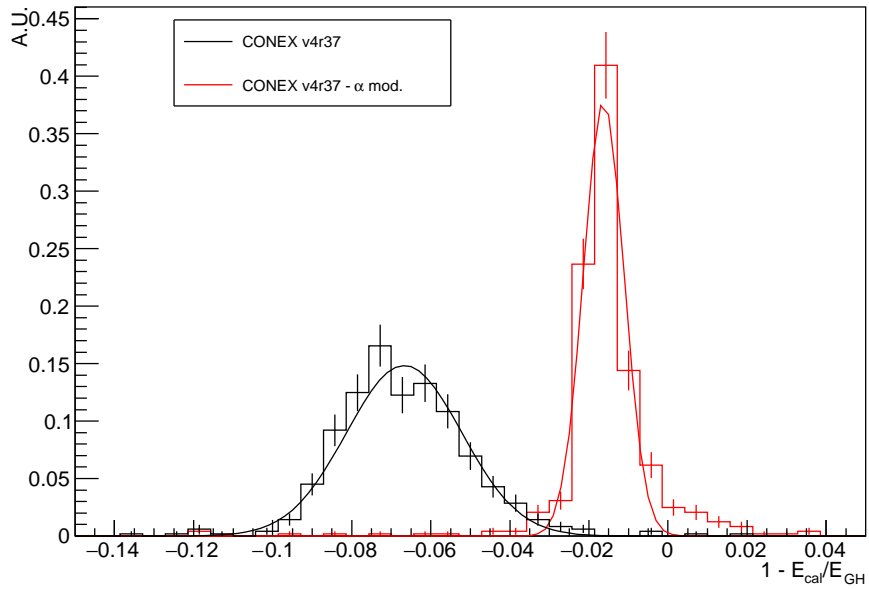}
\end{center}
\caption{
The net effect of $\alpha$ parametrization on reconstructed energy estimate. Results for CONEX modification introduced in the text is shown in red and the original CONEX calculations are depicted in black.}
\label{fig:dE_conex}
\end{figure}

The second cross check was done for showers generated by CONEX.
Following the procedure used for CORSIKA simulations, we present comparison between CORSIKA derived $\alpha$ parametrization and ionization loss rates calculated by the current CONEX v4r37 shown in blue and black in Fig.\ref{fig:scat_conex}, respectively.
A clear discrepancy between above mentioned calculations points to a non--precise description of full MC simulation results by cascade equation approach exploited by CONEX.
This discrepancy is translated into the reconstructed energy estimate depicted in Fig.\ref{fig:dE_conex} by black colour histogram.
A non--negligible bias of $7\%$ together with shower to shower fluctuations of $2\%$ are found.

Fortunately, CONEX software has an intrinsic freedom of parameters that influence the energy deposit calculations \cite{pierog}.
Utilizing this, CONEX v4r37 was modified to match the full MC predicted $\alpha$ parametrization in a better way.
Results of such calculations are shown in Figs.\ref{fig:scat_conex} and \ref{fig:dE_conex} by red colour.
The effect on reconstructed energy is then minimized to the level obtained in the case of CORSIKA full MC simulations.

\section{Conclusions}

We investigated the effect of the mean ionization loss rate parametrization, derived from CORSIKA full MC simulations, to the reconstructed energy estimate.
The estimate is modelled by the integral of the GH function fitted to the longitudinal energy deposit profile of shower.
The $\alpha$ parametrization influences the measurement of light generated proportionally to the $\Nch$, i.e. the events dominated by Cherenkov light.

In the case of CORSIKA full MC simulation, the impact of the $\alpha$ parametrization is at the order of $1-2\%$ depending on low--energy interaction model used.
When the CONEX generated showers are considered, the change in the CONEX code is needed to reduce the $\alpha$ parametrization influence from $7\%$ down to $1\%$ level.
The newly derived CONEX setting will be available in the next CONEX release after more cross checks and improvements are done \cite{pierog}.

\section*{Acknowledgment}
We want to thank Dr. Tanguy Pierog for his kind help in searching for the source of the effective ionisation loss rate discrepancy between CORSIKA and CONEX.
This  work  was  supported  by  the  Czech  Science
Foundation grant 14-17501S.

\end{document}